\documentstyle[12pt]{article}
\def\beq{\begin{equation}}
\def\eeq{\end{equation}}

\begin{document}
\begin{titlepage}
\begin{center}
{\Large \bf Theoretical Physics Institute \\
University of Minnesota \\}  \end{center}
\vspace{0.3in}
\begin{flushright}
TPI-MINN-93/42-T \\
September 1993
\end{flushright}
\vspace{0.4in}
\begin{center}
{\Large \bf Effects of amplitude nullification in the standard
model\\} \vspace{0.2in} {\bf B.H. Smith  \\ }
School of Physics and Astronomy \\
University of Minnesota \\
Minneapolis, MN 55455 \\

\vspace{0.2in}
{\bf   Abstract  \\ }
\end{center}
Delicate cancellations among diagrams can result in the vanishing 
of threshold amplitudes in the standard model.  This phenomenon is 
investigated for multi-Higgs production by scalar, vector, and fermionic 
fields. A surprising gap is found where although the tree-level amplitude 
for $2$ incoming gauge bosons to produce $N_v$ particle may vanish, the 
amplitude for the production of $N_{v}+1$ particles is nonzero.

\end{titlepage}

One of the most interesting pieces of phenomenology to come from the study 
of multi-particle processes is amplitude 
nullification$^{\cite{voloshin3,voloshin4,smith1}}$.  The interest in this 
phenomenon was spurred by an observation about the tree-level 
amplitude for two on-shell scalar particles scattering to produce $n$ 
particles. At the kinematical threshold, this $2 \to 
n$ amplitude vanishes for all $n>4$ in $\lambda\phi^4$ theory with an 
unbroken reflection symmetry$^{\cite{voloshin3}}$.  In $\lambda\phi^{4}$ 
theory with broken symmetry, it was noted that the tree-level threshold 
amplitude of $2 \to n$ production vanishes for all $n>2^{\cite{smith1}}$. 

The nullification phenomenon forbids the threshold amplitude for one 
off-shell scalar decaying into $n$ scalars, $1 \to n$, from developing an 
imaginary part at the one-loop level. At the one-loop level, any relevant 
unitary cut must cross exactly two lines. Any such cut will relate the 
imaginary part of the amplitude to the product of the vanishing $2 \to n$ 
amplitude.

At the threshold, amplitudes of $1 \to n$ processes still exhibit
factorial growth, and, therefore, appear to violate unitarity. It is not 
clear what happens above the threshold, and it may be possible that diagram 
cancellation plays a critical role in the restoration of unitarity. It can 
be speculated that this nullification is the manifestation of some as to yet 
undiscovered symmetry. There exists a scenario where nullification could 
provide a clue as to the restoration unitarity at high multiplicity in the 
standard model$^{\cite{akp3}}$. In order for this to occur, the coupling of 
the Higgs field to the other fields in the standard model would have to take 
on very special values. Requiring these special couplings to occur would 
place restrictions on the ratios of masses in the standard model, and be an 
attractive way of explaining the particle mass spectrum.

For simplicity, we will consider a $U(1)$ gauge theory coupled to a
fermionic field and a complex scalar field. All of the features in the 
following discussion can be easily generalized to the standard model. The 
masses of the gauge and fermionic fields are generated in the standard way 
through the Higgs mechanism. The mass and quadratic coupling of the 
physical scalar particle shall be denoted as $m$ and $\lambda/4$, 
respectively. After symmetry breaking, the scalar field develops a vacuum 
expectation value $v = \sqrt{m^2/(2\lambda)}$. For simplicity, we will 
work in units where $m=1$. The amplitude calculations shall rely on the 
reduction formula technique$^{\cite{brown}}$ for calculating multi-particle 
amplitudes.  The key observation for using this technique, is that at the 
threshold, all fields have zero momenta and are functions of time only. The 
LSZ reduction simplifies to a one dimensional problem. Using this method, 
the amplitude for the threshold decay of one scalar to $n$ scalars 
is,  

\beq 
\langle n|\sigma|0\rangle = \left (
{\partial\over{\partial z}} \right )^{n} 
\langle 0+|\sigma(x)|0- \rangle,
\eeq
where $z(t)$ is the response of the free Higgs field in the presence of
a source, $\rho$, and is proportional to $e^{imt}$, and $\langle
0+|\sigma|0- \rangle$ is the response of the interacting Higgs
field in the presence
of a source. The tree-level amplitudes are generated by approximating
the response by the solution to the classical field equations,
$\sigma_{cl}(t)$. This well known generating function for $1 \to n$
processes is$^{\cite{brown}}$,  
\beq 
\sigma_{cl}(z(t)) =
{z(t)\over{1-z(t)\sqrt{\lambda\over{2 m^{2}}}}}. \label{trexp} 
\eeq

The first loop correction to the amplitude is generated by expanding
the mean value of the field around the classical solution, $\langle
0+|\sigma(x)|0- \rangle = \sigma_{cl}(x)+\sigma_{1}(x)$, where
$\sigma_{1}(x)$ is the mean value of the quantum part of the field. The
equations of motion can be expanded around $\sigma_{cl}(x)$, and only
the leading contribution in $\lambda$ needs to be retained at the one-loop 
level,

\begin{eqnarray}
{\partial^2 \sigma_{1}(t)\over{\partial t}^2} + m^2 \sigma_{1}(t) +6
m\sqrt{\lambda\over{2}}\sigma_{cl}\sigma_{1}(t) +
3\lambda\sigma_{cl}^2\sigma_{1}(t)+  \nonumber \\
V'_{scalar}(t) + V'_{fermion}(t) +
V'_{vector}(t) = 0,
\label{eofm}
\end{eqnarray}

where the three functions, $V'(t)$, represent the
contributions to the amplitude due to scalar, fermion, and vector
field loops. Each of the contributing terms, $V(t)$, in equation 
($\ref{eofm}$) contains a term representing the equal time two-point 
field correlation function evaluated in the $\sigma_{cl}$ background field. 
This can be visualized as the evaluation of a tadpole Feynman diagram, with 
the equal time Green function coming from the propagator in the loop (for a 
discussion on using diagrammatic techniques in the calculation of 
multiparticle processes, see $\cite{smith3}$). 

This two point Green function can be found by taking the inverse of the 
second variation of the Lagrangian evaluated with $\sigma = \sigma_{cl}$. 
The Green function is a transformation of the amplitude of $2$ incoming 
particles to produce $n$ scalars at the threshold. For physical processes, 
the interesting object is the double pole of the amplitude as the momenta 
of the incoming particles are taken to be on mass shell.  These processes can 
be characterized by a single parameter, $\omega = \sqrt{{\bf k}^{2}+1} = 
n/2$, where ${\bf k}$ is the spatial momentum of the incoming particles, 
and $n$ is the number of particles produced in the final state. If the 
Green function does not have a pole for $\omega = n/2$, then the 
tree-level amplitude for $2 \to n$ particles is zero.

In the following sections, we will discuss how particles from each sector of 
the standard model contribute to the one-loop amplitude of $1$ Higgs 
$\to n$ Higgs and where nullification in the production of Higgs 
particles occurs in this sector.

{\bf The Higgs Sector}

Both the one-loop correction to $1 \to n$ processes and $2 \to n$ 
nullfication for incoming scalars have been discussed previously in 
$\cite{smith1}$.  The first variation of the scalar potential 
contributes terms to the equation of motion that 
are both quadratic and cubic in the scalar field, $\sigma$.  Expanding 
around the mean field, 

\beq 
V'_{scalar}(t) = 
3\sqrt{\lambda\over{2}}\langle\sigma_q(x)\sigma_q(x)\rangle  +
3\lambda\sigma_{cl}\langle\sigma_q(x)\sigma_q(x)\rangle,
\label{scalarF}
\eeq
where $\langle\sigma_q(x)\sigma_q(x)\rangle$ is the spatially
independent two-point Green function in the classical background
field, $G(x,x') \equiv \langle T\sigma_q(x)\sigma_q(x')\rangle$, taken
in the limit $x \to x'$. As discussed in a previous work$^{\cite{smith1}}$,
the Green function 
is the inverse of the second variation of the Lagrangian 
evaluated in the classical background field.  

\beq
\left [
\partial^2 + 1 + 6
\sqrt{\lambda\over{2}}\sigma_{cl}(t)+3\lambda\sigma_{cl}(t)^{2} \right ]
G(x,x') = -i\delta(\tau,\tau').
\label{scalarEqn}
\eeq

The component of $G(x,x')$ with spatial momentum ${\bf k}$, denoted as 
$G_{\omega}(\tau,\tau')$ has been constructed from the homogeneous equation 
with the operator in equation ($\ref{scalarEqn}$) 
$^{\cite{smith1}}$. For future reference, we express 
$G_{\omega}(\tau,\tau')$ in terms of a general Green function,

\beq
G_{\omega}(a,\tau,\tau') = 
1/(2\omega)(f^{+}_{a}(\tau)f^{-}_{a}(\tau')\Theta(\tau-\tau') + 
f^{-}_{a}(\tau)f^{+}_{a}(\tau)\Theta(\tau'-\tau)), \eeq 
where $\tau$ is defined by,

\beq
e^{\tau} = -\sqrt{\lambda\over{2}}z(t),
\label{udef}
\eeq
and the two solutions to the homogenous differential equation with the 
operator in equation ($\ref{scalarEqn}$) are

\beq
f^{\pm}_{a} = e^{\mp\omega\tau} F(-a,a+1;1\pm 2\omega;{1\over{1+e^{\tau}}}),
\label{fdef}
\eeq
and $F(a,b;c;z)$ is the hypergeometric function. The hypergeometric series 
in equation ($\ref{fdef}$) will terminate when $a$ is an integer. For 
incoming scalar particles, $G_{\omega}(\tau,\tau') = 
G_{\omega}(2,\tau,\tau')$. If the series did not terminate, the Green 
function would have poles for all half integer values of $\omega$.  Since 
the series terminates, there are poles only for $\omega = \pm 1/2, \pm 1$. 
Thus, the tree-level amplitude for two on-shell scalars to form $n$ scalars 
vanishes at the threshold for $n>2$.

Summing over all momenta modes, the equal time Green function is, aside from 
a renormalization of $m$ and $\lambda$, 

\beq
\langle \sigma_q(x)\sigma_q(x)\rangle ={\sqrt{3}\lambda z(t)^{2} 
\over{4\pi (1-z(t)\sqrt{\lambda/2})^4}}.
\label{greensoln}
\eeq

The Higgs sector cannot contribute an imaginary part to the one-loop $1 \to 
n$ amplitude. There is no two body phase space when $\omega = 1/2, 1$ so it 
is not possible to make a nonzero unitary cut to contribute to the imaginary 
part.

{\bf Fermionic Contribution}

The fermionic contribution to the equation of motion is found by
varying the fermionic portion of the Lagrangian with respect to
$\sigma$. As in the case of the scalar field, the mean value of $\sigma$ is 
expanded around the classical solution, $\sigma_{cl}(t)$. The mean value of 
all other fields is set to zero. The fermionic term in equation 
($\ref{eofm}$) is, 

\beq 
V'_{fermion}(t) = 
{m_{f}\over{v}}\langle\bar{\psi}(x)\psi(x)\rangle, 
\label{fermionF} 
\eeq 
where $\langle\bar{\psi}(x)\psi(x)\rangle$ is the trace two-point
fermionic Green function, $\langle T \bar{\psi}(x)\psi(x')\rangle$, and
$m_{f}$ is the mass of the fermion field as generated by the Higgs
mechanism. The Green function has been discussed in $\cite{voloshin5}$ and 
can be found by calculating the inverse of the variation of the Lagrangian 
with respect to $\psi(x)$ and $\bar{\psi}(x)$ evaluated in the classical
background field, 

\beq 
\left [ i{\not{ \partial}} - m_{f} -
{m_{f}\over{v}}\sigma_{cl}(t) \right ] 
\langle T \bar{\psi}(x)\psi(x')\rangle = -i \delta^{4}(x-x').
\label{fGreenEqn}
\eeq

The spinor structure of equation ($\ref{fGreenEqn}$) is simplified by
the introduction of the Green function $H(x,x')$ defined by,

\beq
\langle T \bar{\psi}(x)\psi(x')\rangle = \left [ -i\not{\partial} - m_{f} -
{m_{f}\over{v}}\sigma_{cl}(t)  \right ] H(x,x'),
\label{Hdef}
\eeq
and analytic continuation into imaginary time as in the previous section.
Since spatial momentum is conserved, the Green
function, $H(x,x')$, can be constructed from the solution to the
homogeneous differential equation,
\beq
\left [ {\partial^2\over{\partial\tau^2}}-\omega^2 + 
\left ( m_{f}^{2} - {m_{f}\over{2}}\gamma_{0} \right )
{1\over{\cosh^{2}{\tau/2}}} \right ] f(\tau) = 0
\label{fhalfgreen}
\eeq

The matrix $\gamma_{0}$ in equation ($\ref{fhalfgreen}$) has
eigenvalues of $+1$ and $-1$. As was noted by Voloshin in
$\cite{voloshin5}$, in order for nullification to take place, the
coefficient of the sech$^{2}(\tau/2)$ term must be $N(N+1)/4$ with
integer $N$ for both eigenvalues. This imposes the conditions on $m_{f}$,

\beq 
4 m_{f}^{2}+2 m_{f} = N_{f}(N_{f}+1),
\label{nfplus}
\eeq

and,

\beq
4 m_{f}^{2}-2 m_{f} = N'_{f}(N'_{f}+1),
\label{nfminus}
\eeq
for some integer values of $N_{f}$ and $N'_{f}$.
Both equation ($\ref{nfplus}$) and equation ($\ref{nfminus}$) are
satisfied when $2 m_{f} = N_{f}$ and $N'_{f} = N_{f}-1$. 

In a basis where the matrix $H$ is diagonal, two of the
diagonal elements of $H$ will be $G_{\omega}(N_{f},\tau,\tau')$ and
two will be $G_{\omega}(N_{f}-1,\tau,\tau')$. The fermionic Green
function for a general fermionic coupling is,

\begin{eqnarray}
\langle\bar{\psi}(x)\psi(x)\rangle = 2
\lim_{\tau' \to \tau} \int{{d^{3}k\over{(2\pi)^3}}}     
\left [ -{\partial\over{d\tau}}G_{\omega}(N_{f},\tau,\tau') + 
{\partial\over{d\tau}}G_{\omega}(N_{f}-1,\tau,\tau') \right ] - 
\nonumber \\ 
2 \lim_{\tau' \to \tau} \int{{d^{3}k\over{(2\pi)^3}}} \left [
\left ( m_{f} + {
m_{f}\over{v}}\sigma_{cl}(t) \right ) \left ( 
G_{\omega}(N_{f},\tau,\tau) +  G_{\omega}(N_{f}-1,\tau,\tau) \right ) 
\right ].  
\label{fermiint} 
\end{eqnarray}
 
Equation ($\ref{fermiint}$) contains two types of terms. One type of
terms contributes only to the renormalization of $m$ and $\lambda$. The
other terms actually contribute to the physical correction to the $1 \to
n$ amplitude. In the theory with $N_{f} = 1$ ($m_{f} = m_{h}/2$), only the
contribution to the renormalization of the coupling constant and mass is
present.

The theory with the simplest nontrivial
$\langle\bar{\psi}(x)\psi(x)\rangle$ and diagram nullification is
characterized by $N_{f} = 2$, i.e. $m_{f} = m_{h}$. In this theory, the
contributions from $G_{\omega}(N_{f}-1,\tau,\tau')$ contribute only to the
renormalization of $m_{h}$ and $\lambda$. Apart from renormalization
terms, the fermionic correlator arising from $G_{\omega}(2,\tau,\tau')$ is,
\beq
\langle\bar{\psi}(x)\psi(x)\rangle = 
-{\sqrt{3}\lambda z(t)^{2} (1+z(t)\sqrt{\lambda/2})
\over{3\pi^{2}(1-z(t)\sqrt{\lambda/2})^5}}.
\eeq
It is worth noting that the fermionic contribution to the equations of
motion and, hence, to the one-loop correction is exactly $-{2\over{3}}$ 
that of the scalar contribution.

{\bf Gauge Field Sector}

The situation with vector fields is more complicated. The 
qualitative behavior of $2$ vector $\to n$ scalar processes is very 
dependant on the polarization of the incoming gauge particles. Care must be 
taken in the choice of gauge. Nonphysical amplitudes like those of the $1 \to
n$ 
processes discussed in previous sections are gauge dependent. While it is 
easiest to calculate in unitary gauge, the gauge loop corrections to 
processes like $1 \to n$ are divergent. These divergences will cancel when a 
gauge independent physical amplitude is calculated. We will work in unitary 
gauge for simplicity of calculation, but we will concentrate on discussing 
the properties of the physical amplitude of two incoming vector bosons 
scattering to form many scalars.


The two point Green function is calculated as in previous sections 
by taking the inverse of the second variation of the Lagrangian. In the 
unitary gauge, the two point vector Green function, $\langle 
V_{\rho}(x)V_{\nu}(x') \rangle$, satisfies the condition, 

\beq 
\left [ 
\partial^{2} + {m_{v}^{2}\over{v^2}} (\sigma_{cl}(t) + v)^{2} \right ] 
\langle V_{\mu}(x)V_{\nu}(x') \rangle - 
\partial^{\mu}\partial^{\rho} \langle V_{\rho}(x)V_{\nu}(x') 
\rangle = -ig_{\mu\nu}\delta^{4}(x-x').  
\label{vectoreqn} 
\eeq 

There are no additional ghost contributions in unitary gauge. It is 
convenient to rotate to a frame of reference where the spatial 
momentum is along the "$1$" axis. The spatially fourier transformed Green 
function,

\beq
G_{\mu\nu}(t,t') = \int \langle V_{\mu}(x)V_{\nu}(x') \rangle e^{i {\bf kx}}
d^{3}{\bf x},
\eeq
has only two nonvanishing derivitives, $\partial_{0}$ and 
$\partial_{1}$. For the transverse components, $G_{22}$ and $G_{33}$, the 
last term on the left hand side of equation ($\ref{vectoreqn}$) does not 
contribute. The Green functions may be written as in previous sections in terms

of the solutions to the homogeneous differential equation constructed from 
the operator in equation ($\ref{vectoreqn}$), 

\beq
G_{\mu\nu} = {1\over{W}}(f_{\mu\nu}^{+}(\tau)f_{\mu\nu}^{-}(\tau') 
\Theta(\tau-\tau') + f_{\mu\nu}^{-}(\tau)f_{\mu\nu}^{+}(\tau') 
\Theta(\tau-\tau')),
\label{gdef} 
\eeq

Where $W$ is the Wronskian of the two solutions. In analogy with previous 
sections, we define the variable, $N_{v}$, which satisfies,

\beq
4 m_{v}^{2} = N_{v}(N_{v}+1).
\eeq

For the the transverse components,

\beq
f_{22}^{\pm} = f_{33}^{\pm} = e^{\omega\tau} F(-N_{v}, N_{v}+1, 1\mp 2\omega
; {1\over{1+e^{\tau}}}).
\eeq
The Wronskian for these solutions is $2\omega$. The transverse part of the 
Green function will have poles at all half integer values of $\omega$ unless
$N_{v}$ is an integer. As noted by Voloshin$^{\cite{voloshin5}}$, the physical
tree-level amplitude for $2$ incoming transversely polarized vector particles
to produce $n$ 
scalars at the threshold vanishes for all $n>N_{v}$. The transverse 
off-diagonal elements of $G_{\mu\nu}$ are zero. 

For longitudinally polarized particles, things are a little bit more 
complicated. Differentiation of equation ($\ref{vectoreqn}$) yields the 
auxillary equation,

\beq
\partial_{\mu} {M^{2}_{v}\tanh^{2}{\tau/2}\over{v^{2}}} G_{\mu\nu} = 
\partial_{\nu}\delta(\tau-\tau').
\label{auxEqn}
\eeq

Searching for solutions of the form,

\beq                                                 
G_{00} (\tau,\tau') = {1\over{\tanh{\tau}\tanh{\tau'}}} H(\tau,\tau') 
-{v^{2}\over{M^{2}_{V}\tanh^{2}{\tau/2}}} \delta(\tau-\tau'), 
\eeq
the Green function, $H(\tau,\tau')$, becomes the object of interest. 
Combining equations ($\ref{auxEqn}$) and ($\ref{vectoreqn}$) with the 
definition of $H(\tau,\tau')$,

\beq
\left [ {d\over{d\tau}^{2}} - \omega^{2} - m_{v}^{2}\tanh^{2}(\tau/2)
- {1\over{2\sinh^{2}{\tau/2}}}
\right ] H(\tau,\tau') = {(\omega^{2}-1)v^{2}\over{M_{v}^{2}}} 
\delta(\tau-\tau').
\label{ptGreen}
\eeq

As in equation ($\ref{gdef}$), $H(\tau)$ can be constructed from the two 
solutions to the homogenous differential equation with the 
operator in equation ($\ref{ptGreen}$). The operator is the same as the
one that appears in the P\"oschl-Teller 
equation. The solution as described in the Appendix is,

\beq
f^{\pm}_{H} = \left [ - 2 {d\over{d\tau}} +a \tanh(\tau/2)+ \coth(\tau/2)
\right ]
e^{\mp\omega\tau} F(-(N_{v}-1), N_{v}; 1\pm 2\omega; {1\over{1+e^{\tau}}}).
\eeq 

     Note that the normalization of the Green function, $H(\tau,\tau')$, 
on the right hand side of equation ($\ref{ptGreen}$) is not unity. As with 
the case of transverse gauge particles, there are poles in $f_{H}$ at every 
half integer value of $\omega$ when $N_{v}$ is not an integer. When $N_{v}$ 
is an integer, there are no poles for $\omega > (N_{v}-1)/2$. The Wronskian 
of the functions $f^{\pm}_{H}$ can be conveniently evaluated at the point 
$\tau = \infty$. The result is

\beq
-\omega \left [ 4\omega^{2}-(N_{v}+1)^2 \right ].
\label{HWronskian}
\eeq

There is an additional pole contributed by the Wronskian at the point 
$\omega = (N_{v}+1)/2$. The unusual pattern of the nullification is the 
following: all process are allowed for $n<N_{v}$. The production of 
$n=N_{v}+1$ (which is forbidden in the transverse mode) is allowed. There is 
a small gap in the allowed region, when $N_{v}$ particles are produced, 
which is not allowed. Prouction of $n$ particles for $n>N_{v}+1$ is forbidden.
The remaining components of $G_{\mu\nu}$ can be easily 
constructed from the above formulae, and have the same poles as $G_{00}$.

Because of the pole at $n=N_{v}+1$, gauge field loops can contribute an 
imaginary part to $1 \to n$ amplitude. The amplitude for the $1 \to n$ 
process is not a gauge invariant object, and is highly divergent in the 
unitary gauge.
 
{\bf Concluding Remarks}

The phenomenology associated with nullification can be quite complex. It's 
presence can prevent the one-loop $1 \to n$ amplitude from developing an 
imaginary part. It can also cause the cancellation of $2 \to n$ diagrams in 
all sectors of the standard model for $n>N$ for some value of $N$. 

The behavior in the longitudinal vectors is quite unexpected. While for 
incoming transverse vectors there is cancellation of $2 \to n$ diagrams 
for all $n>N_{v}$, the process with $n=N_{v}+1$ is allowed for longitudinal 
polarized particles. In addition, the process with $n=N_{v}$ is forbidden 
for longitudinal polarized particles. The pole at $n=N_{v}+1$ is what allows 
the $1 \to n$ amplitude to develop an imaginary part. It is of interest to 
note that the number of poles is the same regardless of the polarization of 
the incoming particles.

Although there exists much knowledge about the phenomenology of 
nullification, little is known about its root cause. It is possible that it 
is associated with some symmetry in the standard model that may play a part 
in the restoration of unitarity at large multiplicities. However, other 
scenerios have been suggested$^{\cite{vg}}$.

I would like to think Mikhail Voloshin for many inciteful discussions. This 
work was supported, in part, by Department of Energy grant 
DOE-DE-AC02-83ER50105.

{\bf Appendix: Solution of the second 
P\"oschl-Teller Equation through the methods of supersymmetric quantum 
mechanics}

In the section on vector fields, the Green function was related to the 
solution to the P\"oschl-Teller equation, which has the form,

\beq
\left [ {d^{2}\over{dx^{2}}} + {a(a+1)\over{\cosh^{2}(x)}} -
{b(b+1)\over{\sinh^{2}(x)}} \right ] f = \lambda f.
\label{pteq}
\eeq
For the Green functions of vector fields, $b$ has the value of 1.

This problem may be elegantly solved by using supersymmetric arguments
(many reviews have been written on the uses of supersymmetry in quantum 
mechanics, e.g. $\cite{gk}$). Consider supersymmetric quantum mechanics 
with two supercharges defined to be, \beq Q_{\pm} = [\pm {d\over{dx}} +W(x)] 
\sigma^{\pm} \label{scdef} \eeq where $\sigma$ denotes the appropriate 2x2 
Pauli spin matrix and the choice of potential is made to be, 

\beq W(x) = a 
\tanh{(x)} + b \coth{(x)} 
\eeq 

The supersymmetric Hamiltonian is the 
anticommutator of the two supercharges, 

\beq 
H = {1\over{2}} \{Q_{+},Q_{-} \} = 
-{d^{2}\over{dx^2}} + (a+b)^{2} - {a\over{\cosh^{2}(x)}} (a-\sigma_{3}) + 
{b\over{\sinh^{2}(x)}} (b-\sigma_{3}).
\eeq

This operator $\sigma_{1}$ had two eigenvalues, $+1$ and $-1$. These
degenerate eigenstates shall be denote as $\psi_{\pm}$, and obey the
eigenvalue equations,

\beq
H^{\pm} \psi^{\pm} \equiv \left [ -{d^{2}\over{dx^2}} + (a+b)^{2} -
{a(a\mp1)\over{\cosh^{2}(x)}}  +  
{b(b\mp1)\over{\sinh^{2}(x)}}  \right ] \psi^{\pm} = E \psi^{\pm}.
\eeq

Consider the case $b=1$. The second term in $H^{+}$ disappears to leave a
familiar operator. The solutions are known to be,
 \beq
\psi^{+}(x) = e^{\pm\omega x} F(-(a-1), a;1\mp\omega; 
{1\over{1+e^{x}}}),
\eeq 
where $\omega = \sqrt{E-(a+1)^{2}}$, and $F(a,b;c;z)$ is the
hypergeometric function. By virtue of the superymmetry, the eigenvalues
of $H^{-}$ are simply $Q_{-}\psi^{+}$. It is easy to see the the
solution to equation ($\ref{pteq}$) for $b=1$ is simply,
\beq
f = \left [ -{d\over{dx}} + a\tanh{(x)} + \coth{(x)}
\right ] e^{\pm\sqrt{\lambda}} F(-(a-1), a;
1\mp\sqrt{\lambda};{1\over{1+e^{x}}}).
\eeq
Solutions to equation ($\ref{pteq}$) for larger integer values of $b$ may
be contructed by iterating this procedure.

\end{document}